\documentclass[a4paper,12pt,numbers,sort&compress]{article}

\usepackage{amsmath,amssymb,amsfonts,epsfig,cite,setspace,rotating}

\pdfoutput=1

\begin{document}


\vskip 0.25in

\newcommand{\todo}[1]{{\bf ?????!!!! #1 ?????!!!!}\marginpar{$\Longleftarrow$}}
\newcommand{\fref}[1]{Figure~\ref{#1}}
\newcommand{\sref}[1]{\S\ref{#1}}

\newcommand{\nn}{\nonumber}
\newcommand{\tr}{\mathop{\rm Tr}}
\newcommand{\firr}[1]{{}^{{\rm Irr}}\!{\cal F}^{\flat}_{#1}}

\newcommand{\comment}[1]{}

\newcommand{\cM}{{\cal M}}
\newcommand{\cW}{{\cal W}}
\newcommand{\cN}{{\cal N}}
\newcommand{\cZ}{{\cal Z}}
\newcommand{\cO}{{\cal O}}
\newcommand{\cB}{{\cal B}}
\newcommand{\cC}{{\cal C}}
\newcommand{\cD}{{\cal D}}
\newcommand{\cE}{{\cal E}}
\newcommand{\cF}{{\cal F}}
\newcommand{\cX}{{\cal X}}
\newcommand{\IA}{\mathbb{A}}
\newcommand{\IP}{\mathbb{P}}
\newcommand{\IQ}{\mathbb{Q}}
\newcommand{\IR}{\mathbb{R}}
\newcommand{\IC}{\mathbb{C}}
\newcommand{\IF}{\mathbb{F}}
\newcommand{\IV}{\mathbb{V}}
\newcommand{\II}{\mathbb{I}}
\newcommand{\IZ}{\mathbb{Z}}
\newcommand{\re}{{\rm Re}}
\newcommand{\im}{{\rm Im}}

\newcommand{\CA}{\mathbb A}
\newcommand{\CP}{\mathbb P}
\newcommand{\tmat}[1]{{\tiny \left(\begin{matrix} #1 \end{matrix}\right)}}

\newcommand{\diff}[2]{\frac{\partial #1}{\partial #2}}
\newcommand{\diag}{{\rm Diag}}

\newcommand{\drawsquare}[2]{\hbox{%
\rule{#2pt}{#1pt}\hskip-#2pt
\rule{#1pt}{#2pt}\hskip-#1pt
\rule[#1pt]{#1pt}{#2pt}}\rule[#1pt]{#2pt}{#2pt}\hskip-#2pt
\rule{#2pt}{#1pt}}
\newcommand{\fund}{\raisebox{-.5pt}{\drawsquare{6.5}{0.4}}}
\newcommand{\antifund}{\overline{\fund}}

\newtheorem{theorem}{\bf THEOREM}
\def\thetheorem{\thesection.\arabic{theorem}}
\newtheorem{conjecture}{\bf CONJECTURE}
\def\thetheorem{\thesection.\arabic{conjecture}}
\newtheorem{observation}{\bf OBSERVATION}
\def\thetheorem{\thesection.\arabic{observation}}

\def\theequation{\thesection.\arabic{equation}}
\newcommand{\setall}{\setcounter{equation}{0}
        \setcounter{theorem}{0}}
\newcommand{\setequation}{\setcounter{equation}{0}}

\renewcommand{\thefootnote}{\fnsymbol{footnote}}
\centerline{\Huge Graph Zeta Function and Gauge Theories}
~\\
\vskip 2mm
\centerline{
{\large Yang-Hui He}\footnote{\tt Yang-Hui.He.1@city.ac.uk, hey@maths.ox.ac.uk}
}
~\\
{\scriptsize
\begin{center}
\begin{tabular}{ll}
  $^1$ & {\it Department of Mathematics, City University,} 
  {\it Northampton Square, London EC1V 0HB, U.K.}\\
  $^2$ & {\it School of Physics, NanKai University,}
  {\it Tianjin, 300071, P.R.~China}\\
  $^3$ & {\it 
       Merton College, Oxford, OX1 4JD, U.K.}\\
\end{tabular}
\end{center}
}
~\\
~\\

\begin{abstract}
Along the recently trodden path of studying certain number theoretic properties of gauge theories, especially supersymmetric theories whose vacuum manifolds are non-trivial, we investigate Ihara's Graph Zeta Function for large classes of quiver theories and periodic tilings by bi-partite graphs.
In particular, we examine issues such as the spectra of the adjacency and whether the gauge theory satisfies the strong and weak versions of the graph theoretical analogue of the Riemann Hypothesis.
\end{abstract}

\newpage
\tableofcontents

\newpage

\section{Introduction}\setall
Though the geometry of gauge theories has constituted one of the greatest intellectual triumphs and served as a heart-warming matrimony between physics and mathematics, the infusion of number theory, that Queen of mathematics, into or her emergence from physics has been comparatively humble\footnote{
Quoting P.~Candelas, FRS, from his recent London Triangle talk, this mysterious connection, which has already produced some beautiful results, beckons for deeper exploration.
}.
Yet, there has been a colourful skein threading the two seemingly disparate enterprises since at least Hilbert and Polya in addressing the Queen of number theory, the Riemann Hypothesis: how one might find a Hermitian operator whose eigenvalues coincide with the critical zeros of the Riemann zeta function.

Over the decades, there have been various proposals in finding analogues of the zeta function which satisfy a functional equation symmetric about the vertical line situated at $\frac12$ in the complex plane, as well as possess suggestive zero free regions.
The most remarkable of these is indubitably the Hasse-Weil zeta function which is an (exponentiated) generating function for algebraic varieties over finite number fields.
The relation of its functional form to the geometrical data has been one of the greatest feats in modern mathematics.
Recently, there has been some activity in investigating the physics of this zeta function, for compact Calabi-Yau manifolds \cite{Candelas:2000fq}, as well as for non-compact vacuum varieties in relation to gauge theory \cite{He:2010mh}.

This latter case of supersymmetric gauge theories with non-trivial vacuum promises a particularly enticing ground for the inter-play amongst physics, algebraic geometry, combinatorics and number theory.
Physically, they are generically theories with product gauge groups and comprise of such important examples as the Standard Model and those arising from the AdS/CFT Correspondence. A so-called plethystic programme \cite{Benvenuti:2006qr}, exploiting algebro-geometric and combinatorial techniques, has been developed to enumerate the gauge invariants in such theories, as too was the goal of \cite{He:2010mh} to relate the relevant generating functions to the Hasse-Weil zeta function of the vacuum moduli space.

One salient and, as it transpires, important feature of these supersymmetric gauge theories is that they have an underlying structure of a finite, labeled, directed graph with a path-algebra encoding the interaction terms in the Lagrangian. Thus was germinated the industry of quiver gauge theories \cite{Douglas:1996sw}. In fact, for a vast class of these theories, {\it viz.}, those whose vacuum affords a toric description \cite{Feng:2000mi} - consisting of the majority of the theories in the AdS/CFT correspondence, as the Calabi-Yau cones involved are largely the only ones whose metrics are known explicitly - a simpler still graphical description exists.
For the myriad of these theories, infinite families thereof having already been identified, a bi-partite graph (dimer model) on a torus, or, equivalently called a periodic brane tiling of the plane, suffices completely to encode the physics \cite{Kennaway:2007tq,Yamazaki:2008bt}.

The study of these graphs has been a vast subject over the past decade and the algebraic and differential geometry of the associated Calabi-Yau space, as well as the combinatorics emergent from enumerations of inequivalent theories and counting of BPS spectra, have been exploited to remarkable and beautiful consequences.
It is not the intention of the present writing to address these issues.
It is, however, our purpose to investigate, in the spirit of \cite{He:2010mh}, the number theoretical properties which underly the construction.
Indeed, in {\it loc.~cit.}, given that the central object is an algebraic variety, it was natural to study these affine varieties over finite fields.
Here, as another central object is a finite graph, it would be most conducive to examine these theories under the number theoretical guise of these graphs.

Luckily, the concept of a zeta function for a finite graph, in the fine tradition of Weil {\it et al.}, has been long extant: this is the Ihara-Serre zeta function.
The subject is vast and has been generalized over the decades to encompass all partially directed finite graphs.
The analogues of primes, functional equations, eigen-spectrum and Riemann Hypothesis have all been established or conjectured.

Thus a natural course of action recommends itself and we shall allow ourselves to be freely lead by this path.
In \S\ref{s:ihara}, we familiarize ourselves with the rudiments of the Ihara zeta function and set the nomenclature which will facilitate explicit evaluation.
Next, in \S\ref{s:quiver}, we apply the techniques to the non-chiral quiver theories whose vacuum moduli spaces are affine Calabi-Yau varieties.
We then proceed to address the bi-partite graphs which represent gauge theories which inherently have a toric geometry in \S\ref{s:dimer}.
Finally, we conclude with prospects in \S\ref{s:conc}.

\section{The Graph Zeta Function}\label{s:ihara}
In his classification of the discrete subgroups of $PGL(2; k)$, the projective general linear group in dimension 2 over a p-adic number field $k$, Ihara \cite{ihara} constructed a zeta function of Selberg type; {\it i.e.}, a zeta function whose analytical structure is determined by the spectral data of Laplace operators on a surface.
Serre pointed out some years thereafter that Ihara's construction had an underlying graphic interpretation \cite{serre-tree} and thus was born the concept of the {\bf graph zeta function}.

In particular, the Ihara-Serre graph is an undirected one and over the years \cite{Hashimoto,Horton,M-S,Sato,S-T} there have been various generalizations and extensions, culminating in the recent paper \cite{T-P} which presents a succinct explicit formula for the zeta-function of any finite graph.
The reader is referred to a thorough introduction to the subject in \cite{Terras} as well as some charming accounts in various lecture notes \cite{Terras-notes}; from these we first take some requisite rudiments and develop an appreciation before moving onto the aforementioned general expressions recently obtained.

The initial definition of the Ihara zeta function was for graphs $G$ which are finite ({\it i.e.}, having a finite number of edges and nodes, say $m$ and $n$, respectively), {\em undirected} ({\it i.e.}, the edges have no direction), connected ({\it i.e.}, each node can be reached by traversing some edge) and usually tail-less by containing no degree-1 vertices ({\it i.e.}, no node which is connected by a single edge only; recall that the {\bf degree} (or sometimes called {\bf valency}) of a node is the number of vertices adjacent thereto by the connection of an edge).
These restrictions will be relaxed as we proceed.
The graph was, however, allowed to have multiple edges and loops.

The $n \times n$ {\bf adjacency matrix} is such that the $(i,j)$-th entry is the number of edges from node $i$ to $j$, taken with the convention that the $i$-th diagonal entry is twice the number of self-adjoining loops to the $i$-th node.
We can further assign a standard direction in $G$ by making each of the $m$ undirected edges a bi-directional arrow and introduce a label $e_1, \ldots, e_m$ for one set of the arrows and $e_{m+1} = e_1^{-1}, \ldots, e_{2m} = e_m^{-1}$ the contrary set in the oppose direction.
A {\bf primitive path} $P$ in $G$ is then a closed path which never back-tracks or has tails. 
This is to say that in the path along directed edges $e_{i_1} \ldots e_{i_s}$, beginning and terminating on the same node, no $e_{j}$ is ever equal to $e_{j+1}^{-1}$ nor can the first arrow be the back-track of the last: $e_{i_1} \ne e_{i_s}$.
Furthermore, we define two closed paths to be equivalent if they are identical but for the choice of the initial (and thus terminal) node; the {\bf length} $\ell(P)$ of the path $P$ is, naturally, the number of edges traversed.
At last, we can define the most useful concept for a zeta-function: a {\bf prime} in $G$ is the equivalence class of prime paths.

\begin{figure}[!h!t!b]
\centerline{
\includegraphics[trim=0mm 0mm 0mm 0mm, clip, width=5.5in]{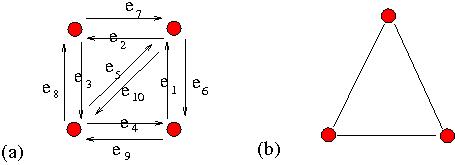}
}
\caption{{\sf 
Examples of finite graphs: (a) a non-chiral quiver whose every edge has been replaced by a bi-directional arrow;
(b) the Cayley graph for $\IZ_3$.
In our nomenclature, the former is a fully directed graph, and the latter, fully undirected. A partially directed graph would consist of a mixture of edges and arrows.
}
\label{f:eg-path}}
\end{figure}

It is perhaps helpful to illustrate these concepts diagrammatically and we shall take a figure from \cite{Terras} (which, for amusement, should be reminiscent of a well-known quiver for Hirzebruch, though here undirected) and reproduce in part (a) of \fref{f:eg-path}.
The paths $[e_2e_3e_5]$, $[e_1e_2e_3e_4]$, and $[e_1e_2e_3e_4e_1e_{10}e_4]$, of lengths respectively 3, 4, and 7, are clearly prime, but so is the infinite family $[(e_1e_2e_3e_4)^ne_1e_{10}e_4]$ for any $n \in \IZ_{\ge 1}$.
It must be warned that in this path algebra, we do not have unique factorization into primes.

Thus prepared, the Ihara or {\bf Graph Zeta function} for a tail-less finite undirected graph $G$ is defined to be the formal (possibly) infinite product over primes and in a complex variable $z$:
\begin{equation}
\zeta_G(z) := \prod\limits_{[P] \mbox{ prime}} 
  \left(1-z^{\ell([P])}\right)^{-1} \ ,
\end{equation}
where, we recall, $\ell([P])$ is the length of the path.
Moreover, in the product over primes, we distinguish between $[P]$ and $[P^{-1}]$ since orientation reversal is not in the same equivalence class, which is reserved, as stated above, only for cyclic relabeling of the edges traversed.
The similarity to and inspiration drawn from the famous Euler product $\prod\limits_{p \mbox{ prime}} \left(1 - p^{-s}\right)^{-1}$ for the Riemann zeta function $\zeta(s)$ is obvious.
The depth of this analogy, of course, transcends a mere formal resemblance and the number-theoretical groundings have supplanted the motivations for many investigations.
Some salient results have included the following.

First, let us give some definitions.
We call a $(q+1)$-regular graph $G$ ({\it i.e.}, each and every node has degree $q+1$, and hence $q+1$ adjacent neighbours) {\bf Ramanujan} if the maximum of the absolute value of eigenvalues of the adjacency matrix $A$, excluding $q+1$ itself, is bounded by $2 \sqrt{q}$.
In other words, $\max\left\{ |\lambda| : \lambda \in \mbox{Spec}(A), |\lambda| \ne q+1 \right\} \le 2 \sqrt{q}$.
Moreover, the Ihara zeta function of a $(q+1)$-regular (connected, tail-less) graph is said to satisfy the {\bf Riemann Hypothesis} if for $0 < \re(s) < 1$, the zeros of $\zeta_G(q^{-s})^{-1}$ lie on the $\re(s) = \frac12$ line.
Armed with such suggestive terminology we have the following known results enjoyed by the zeta function for a finite graph of $n$ nodes and $m$ edges without tails:
\begin{itemize}
\item Analogue of Weil Conjectures: $\zeta_G(z)$ is rational and is, in fact, the inverse of a polynomial.
\item Analogue of the Prime Number Theorem: defining $\pi(m)$ as the number of primes in the graph $G$ of length $m$, $\Delta_G$ as the greatest common divisor of the lengths of primes, $R_X$ the radius of convergence of $\zeta_G(z)$ ({\it i.e.}, the position of the nearest pole to the origin), then $\pi(m) = 0$ if $\Delta_G$ does not divide $m$, otherwise, $\pi(m) \sim \frac{\Delta_G}{m R_G^m}$ as $m$ tends to infinity.
\item There is a {\em functional equation} for the graph xi-function for $(q+1)$-regular $G$, defined as
  $\xi_G(z) := (1+z)^{m-n}(1-z)^m(1-q z)^n \zeta_G(z)$, {\it viz.},
  $\xi_G(z) = \xi_G(\frac{1}{qz})$.
\item For $(q+1)$-regular $G$, $\zeta_G(z)$ satisfies the Riemann Hypothesis iff $G$ is Ramanujan.
\end{itemize}

These above elegant properties are indeed inviting.
What is perhaps intimidating is the seeming abstruseness in the definition of the Ihara zeta function and a desire for a computationally feasible expression has emerged since the original works.
Over the years, the various authors in {\it cit.~ibid.} have been writing explicit formulae for various generalizations of the class of graphs encountered above.
We resort to a recent work of \cite{T-P} which gives a closed form, or a working definition if you will, of the Ihara zeta function for an {\em arbitrary finite graph} $G$:
\begin{equation}\label{ihara}
\zeta_G(z) = \frac{(1-z^2)^{-\tr(Q-I)/2}}{\det(I - A z + Q z^2 + P z^3)} \ ;
\end{equation}
this shall be the expression which we shall henceforth employ.
A few clarifications on the notations above:
\begin{itemize}
\item The graph $G$, with $n$ nodes, is allowed to have both edges (undirected) and arrows (directed); such a graph is called {\em partially directed}.
All subsequent matrices are $n \times n$.
\item The matrix $A$ is the {\em undirected} adjacency matrix, whose $(i,j)$-th entry is the number of length-one walks, along either edges or arrows, from node $i$ to node $j$.
We emphasize that the convention is to have the diagonal entries of $A$ being {\em twice} the number of self-adjoining loops.
\item The matrix $P$ is the {\em directed} adjacency matrix\footnote{
This matrix distinguishes, for example, the situation where a node has (1) an edge loop or (2)two arrow loops, both of which have the same $A$ (note that they are both length 1 because the two arrows are in the same equivalence class).
In our convention, however, all self-adjoint loops and bi-directional arrows will be denoted by edges.
} whose $(i,j)$-th entry is the number of arrows from node $i$ to node $j$.
Thus, when $G$ is fully undirected (having no arrows) $P$ is the zero matrix and when it is fully directed (having no edges) $P=A$.
\item The matrix $Q$ is the matrix of undirected degrees, defined as follows.
For a node $i$, the undirected degree is the number of walks of length 1 starting at $i$ using only (undirected) edges. That is to say, it is the number of adjacent nodes (including itself) by edges only and not by arrows. $Q$ is then the diagonal matrix of the undirected degrees of the vertices subtracted, by convention, by the identity matrix.
For any graph, one can readily see that the exponent of the numerator $-\tr(Q-I)/2$ is equal to the number of nodes minus the number of edges. Note that for any self-adjoining loops, the undirected degree is counted as 2, as is with the matrix $A$, because the beginning and the end of the edge is ambiguous.
\end{itemize}

For illustrative purposes, let us give a hypothetical example with the appropriate matrices indicated
\[
\begin{array}{lll}
\includegraphics[trim=0mm 0mm 0mm 0mm, clip, width=2in]{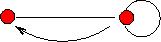}
&&
A = \left(\begin{array}{cc} 0 & 1 \\ 2 & 2 \end{array} \right) , \
P = \left(\begin{array}{cc} 0 & 0 \\ 1 & 0 \end{array} \right) , \
Q = \left(\begin{array}{cc} 1 & 0 \\ 0 & 1+2 \end{array} \right) - 
    I_{2 \times 2}
    \ .
\end{array}
\]
We see that $A$ gives the full adjacency information of arrows and edges while $P$ encodes only the direct arrows.
Moreover, $-\tr(Q-I)/2 = -\frac12 \tr
\left(\begin{array}{cc} -1 & 0 \\ 0 & 1 \end{array} \right) = 0$ as indeed there are two nodes with 2 edges (the other being an arrow) so that their difference is 0.

Once we have the form of the Ihara zeta function explicitly, we can use the primary definition of the graph zeta function as a generating function for enumerating the primes, {\it i.e.}, minimal non-back-tracking (Eulerian) cycles.
This is done by taking the logarithmic derivative and series expanding $z \frac{d}{dz} \log \zeta_G(z)$: the $n$-th coefficient will give the number of length-$n$ prime paths.
Thus concludes our introduction to the graph zeta function and in the next section we will move onto an intriguing category of graphs to which we will happily apply our new acquired technology.

\section{Quiver Gauge Theories}\label{s:quiver}
Quiver gauge theories have over the past few decades attracted substantial interest both in physics and in mathematics, furnishing, in the former, large classes of gauge theories with product gauge groups, of potential phenomenological interest, and, in the latter, representations of finite graphs, especially in the context of algebraic geometry.
The constant dialogue between the two has been most fruitful and we refer the reader, for example, to \cite{He:1999xj,He:2004rn} for an elementary introduction as well as a recent thesis for a nice comprehensive discussion \cite{Forcella:2009bv}.

To set nomenclature, our quivers will be finite, labeled, directed graphs, allowing for loops, bi-directional arrows and self-adjoining vertices.
The corresponding (super-symmetric) gauge theory will be taken to have gauge group $\prod SU(N_i)$ where $N_i$ is the integer labeling of the $i$-th node.
The matter content consists of bi-fundamentals so that each arrow from node $i$ to $j$ is a field transforming under the $(N_i, \overline{N_j})$ of the factor $SU(N_i) \times SU(N_j)$ in the gauge group.
To fully explore the Ihara zeta function, we will adopt the convention that bi-directional arrows as well as arrows which join a node to itself will be replaced by an edge so that our quivers could be partially directed.
A gauge theory which consists only of edges is called a {\em non-chiral} theory.
In representation theory, the vector space $\IC^{N_i}$ will be associated to the $i$-th node and the arrows are maps in $\hom(\IC^{N_i}, \IC^{N_j})$.
Furthermore, the arrows could obey some (non-commutative) relations, whereby generating a path algebra; in physics such constraints come from a superpotential which governs the interactions.
Let us not, for the present purposes of the graph zeta function, dwell upon these; we will return to this point in \S\ref{s:dimer}.

\subsection{ADE Quivers and $\cN=2$ Gauge Theories}\label{s:ade}
\begin{figure}[!h!t!b]
\centerline{
\includegraphics[trim=0mm 0mm 0mm 0mm, clip, width=6.0in]{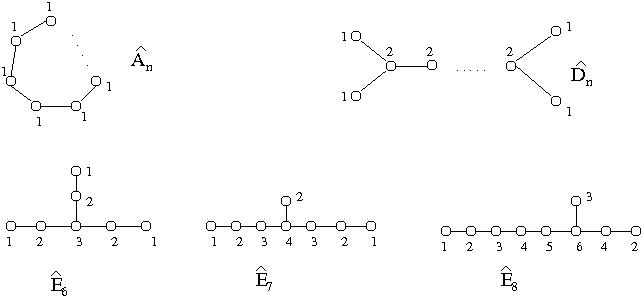}
}
\caption{{\sf 
The ADE Quivers, {\it i.e.}, the simply-laced Dynkin diagrams of affine Lie algebras.
The nodes are labeled by dual Coxeter numbers, or dimensions of the irreducible representations of the corresponding discrete finite subgroup of $SU(2)$.  
}
\label{f:ade}}
\end{figure}

A natural starting point of our investigations is the ADE quivers which are the Dynkin diagrams of the simply laced affine Lie groups.
We recall these in \fref{f:ade}.
As gauge theories, these have gauge groups corresponding to 4-dimensional theories with $\cN=2$ supersymmetry and the dual Coxeter labels give the $SU(N_i)$ factors.
It turns out that there is enough supersymmetry here to complete specify the interactions so we need not worry about the superpotentials in any case.

The famous correspondence of J.~McKay \cite{mckay} states that if one took the fundamental two-dimensional defining representation ${\bf 2}$ of the finite discrete subgroups $\Gamma$ of $SU(2)$ -  whose classification dates back at least to Klein and which we present below in \eqref{ade} to remind the reader -  and formed the Clebsch-Gordan (multiplicity) coefficients $a_{ij}$ with all the irreducible representations ${\bf r}$ of $\Gamma$, as in ${\bf 2}\otimes {\bf r}_i=\bigoplus\limits_{j}a_{ij}{\bf r}_j$, then $a_{ij}$ is the (undirected) adjacency matrix of the graphs in \fref{f:ade}.

There are beautiful algebro-geometric interpretations of this correspondence wherein one looks at the quotient singularity $\IC^2/\Gamma$, all of which can be written as a hypersurface affine variety in $\IC^3$, whose forms we also present in \eqref{ade}. Then, the intersection matrix of the exceptional $\IP^1$-divisors in the Gorenstein resolution of the singularity is exactly the aforementioned $a_{ij}$.
\begin{equation}\label{ade}
\begin{array}{|c|l|c|}
\hline
\mbox{Group} & \mbox{Name} & \mbox{Order} \\ \hline
A_n \simeq \IZ_{n+1} & \mbox{Cyclic} & n+1 \\ \hline
D_n & \mbox{Binary Dihedral} & 2n \\ \hline
E_6 & \mbox{Binary Tetrahedral} & 24 \\ \hline
E_7 & \mbox{Binary Octahedral (Cubic)} & 48 \\ \hline
E_8 & \mbox{Binary Icosahedral (Dodecadedral)} & 120 \\ \hline
\end{array}
\quad
\begin{array}{|ll|}\hline
\mbox{Singularity} & \mbox{Hypersurface} \\ \hline
A_n: &xy+z^{n-1}=0\\
D_n: &x^2+y^2z+z^{n-1}=0\\
E_6: &x^2+y^3+z^4=0\\
E_7: &x^2+y^3+yz^3=0\\
E_8: &x^2+y^3+z^5=0 \\\hline
\end{array}
\end{equation}

Because $\IC^2/ (\Gamma \subset SU(2))$ are locally Ricci flat and resolve to Calabi-Yau two-folds ({\it i.e.}, K3-surfaces), the geometrical interpretation lends itself well to string theory.
Indeed, the world volume theory of a D3-brane transverse to the local Calabi-Yau threefold $\IC \times ADE$ is precisely the $\cN=2$ quiver gauge theory described above\footnote{Strictly, because of the extra factor of $\IC$, to each node in the quiver, there is a doublet of self-adjoining loops. We shall consider this situation shortly.
}.

\subsubsection{The Cyclic Series}
Let us compute the Ihara zeta function for the ADE graphs, which are all undirected and thus are composed only of edges.
As a warm-up, consider the $A_2 = \IZ_3$ case, the quiver is cycle with 3 nodes each of which is connected to the other two by an edge.
This is, in fact, precisely the Cayley graph for $\IZ_3$, as drawn in part (b) 
Therefore, $A = {\tiny \left(
\begin{array}{ccc}
 0 & 1 & 1 \\
 1 & 0 & 1 \\
 1 & 1 & 0
\end{array}
\right)}$, $P=0_{3 \times 3}$ and $Q=I_{3 \times 3}$, giving us, by \eqref{ihara}, $\zeta_{A_2}(z) = (1 - 2 z^3 + z^6)^{-1}$, which is indeed the reciprocal of a polynomial, as required by the graph analogue of the Weil Conjectures.

We can proceed in general and see that $Q$ is always the identity and $P$, the zero matrix and $A$, the circulant matrix with near-diagonal 1's, whence:
\begin{equation}\label{An-noloop}
\zeta_{A_n}(z)^{-1} = \left\{
\begin{array}{ll}
1 - 2z + z^2 & n = 0 \\
1 & n = 1 \\
1 - 2 z^{n+1} + z^{2n+2} & n \ge 2 \ ;
\end{array}
\right.
\end{equation}
the degenerate case of $n=0$ is the trivial 1-noded quiver with a single self-adjoining loop, for which we need $A = Q = \{\{ 2 \}\}\ , P = \{\{0\}\}$, giving us $\zeta_{A_0}(z)^{-1} = (1-z)^2$. 
This geometrically is the flat space $\IC^2$.
For $n=1$, we have the Cayley graph for $\IZ_2$ which, of course, has no loops, and hence the zeta function is just the identity.

Indeed, these are all $2$-regular graphs, with $q=1$, $m=n$.
Recalling our discussions above, we can define the graph xi-function as 
$\xi_{A_n}(z) = (1-z)^{2(n+1)} \zeta_{A_n}$.
For $n\ge 2$, $\xi_{A_n}(z)$ is simply 1 and trivially satisfies the functional equation and indeed the eigenvalues of $A$, being those of the sum of two roots of unity, are bounded by $2\sqrt{q} = 2$, and hence satisfy the Ramanujan property.
For $n=0$, $\xi_{\IC^2}(z)$ is also equal to unity, as is the quiver Ramanujan. The situation of $n=1$ is degenerate since here $q=0$ and there is no functional equation to speak of, as too is the Ramanujan property violated.

Parenthetically, had we adopted the convention of making each (undirected) edge a bi-directional arrow, then we would have $P = A$ and $Q = -I_{n \times n}$.
This gives us $\zeta_{A_{n-1}} = (1-z^2)^n (\det(1-Az)(1-z^2))^{-1} = (\det(1-Az))^{-1}$, which is, in fact, the general result for all fully directed graphs \cite{M-S}.

\paragraph{Self-Adjoining Loops: } 
we mentioned that one should really add two self-adjoining loop to each node in the case of 4-dimensional $\cN=2$ gauge theory, in which case for undirected edges, we would simply modify $A  \to A + 2 \times 2 I$.
The first factor of two is by convention and the second is natural both physically and mathematically, furnishing, in the former, the trivial matter content arising from the ${\bf 1}^{\oplus 2}$ when decomposing the fermionic ${\bf 4} \to {\bf 1}^{\oplus 2} \oplus {\bf 2} $ (or bosonic ${\bf 6} \to {\bf 1}^{\oplus 2} \oplus {\bf 2}^{\oplus 2}$ ) from the parent $\cN=4$ theory upon the orbifolding \cite{Hanany:1998sd,He:1999xj}, and in the latter, the $-2$ self-intersection of the exceptional divisors in the blowup. From a representation theoretical view-point, the new $A$ is now $2 I$ subtracted by the Cartan matrix of the corresponding affine Lie algebra.

Such a modification to the diagonal of $A$ brings about a much more complicated form for the Ihara zeta function.
A general form is difficult to write down, but we here present some first values:
\begin{equation}
\begin{array}{c|l}
n & \zeta_{A_{n}}(z)^{-1} \\ \hline
0 & \left(z^2-1\right)^2 \left(5 z^2-6 z+1\right)\\
1 & (z-1)^4 (z+1)^4 \left(5 z^2-5 z+1\right) \left(5 z^2-3 z+1\right) \\
2 & (z-1)^7 (z+1)^6 (5 z-1) \left(5 z^2-3 z+1\right)^2 \\
3 & (z-1)^9 (z+1)^8 (5 z-1) \left(5 z^2-4 z+1\right)^2 \left(5 z^2-2 z+1\right)
\\
4 & (z-1)^{11} (z+1)^{10} (5 z-1) \left(25 z^4-35 z^3+21 z^2-7 z+1\right)^2
\\
5 & (z-1)^{13} (z+1)^{12} (5 z-1) \left(5 z^2-5 z+1\right)^2 \left(5 z^2-3 z+1\right)^2 \left(5 z^2-2 z+1\right)
\end{array}
\end{equation}
We note that the addition of the adjoints makes these quivers $4$-regular graphs, so that $q=3$.
However, the Ramanujan condition is violated in general because we can see explicitly the existence of eigenvalues between the trivial eigenvalue of 4 and the bound $2\sqrt{3}$, for sufficiently large $n$.
\comment{
Indeed, this is reflected by the fact that the functional equation, due to the palindromicity, satisfied by the zeta function is the rather unexciting $\zeta_{A_n}(\frac{1}{z}) = z^n \zeta_{A_n}(z)$, instead of one which could allow for the defining of a xi-function symmetric about the critical line of $\re(z) = \frac12$.}
\subsubsection{The Dihedral Series}
Let us now depart the comfortable realm of regular graphs and proceed on to the D and E series.
With zeros on the diagonals of the $A$-matrix, we readily find the trivial result that for all $n$, $\zeta_{D_{n}}(z) = 1$.
This is immediate from the fact that there are no cycles in the quiver.
After all, it is important to have cycles in order to have gauge invariant operators (cf.~\cite{Benvenuti:2006qr,Hewlett:2009bx,deMedeiros:2010pr}).
Therefore, we shall include the double self-adjoining loops as discussed above by letting the diagonal entries of $A$ be 4, generating much more involved expressions, some incipients of which are:
\begin{equation}
\begin{array}{l|l}
n & \zeta_{D_{n}}(z)^{-1} \\ \hline
1 & -(z-1)^{10} (z+1)^9 (2 z-1)^6 \left(28 z^3-16 z^2+7 z-1\right)\\
2 & -(z-1)^{12} (z+1)^{11} (2 z-1)^4 \left(24 z^3-20 z^2+8 z-1\right) \left(24 z^4-36 z^3+20 z^2-7 z+1\right)\\
3 & -(z-1)^{14} (z+1)^{13} (2 z-1)^4 \left(24 z^4-40 z^3+24 z^2-8 z+1\right) \left(120 z^5- \right. \\
  & \left. \qquad \qquad 176 z^4+120 z^3-48 z^2+11 z-1\right)\\
4 & -(z-1)^{16} (z+1)^{15} (2 z-1)^4 \left(120 z^5-200 z^4+140 z^3-56 z^2+12 z-1\right)
   \left(120 z^6 \right. \\
   & \left. \qquad \qquad -272 z^5+260 z^4-148 z^3+52 z^2-11 z+1\right)\\
5 & -(z-1)^{18} (z+1)^{17} (2 z-1)^4 \left(120 z^6-296 z^5+300 z^4-172 z^3+60 z^2-12
   z+1\right) \left(600 z^7 \right. \\
   & \left. \qquad \qquad 
   -1360 z^6+1396 z^5-880 z^4+360 z^3-96 z^2+15 z-1\right)
\end{array}
\end{equation}

What can one say about the spectral as well as analytic properties?
There are some partial results for irregular graphs into whose territory we have now entered; one useful theorem, due to Kotani-Sunada (cf.~Chapter 8 of \cite{Terras}), is that the radius of convergence $R_G$ of $\zeta_G(z)$ (which we recall to be the position of the pole nearest to the origin) lies between the reciprocals of the minimal degree $p+1$ and maximal degree $q+1$ in $G$:
\begin{equation}\label{pq}
q^{-1} \le R_G \le p^{-1} \ .
\end{equation}
We see that the radius is rather small indeed, never exceeding unity.
In fact, their theorem further guarantees that every pole of $\zeta_G(z)$ has absolute value between $R_G$ and 1, inclusive.

On these grounds, the {\bf graph theory Riemann Hypothesis} is that $\zeta_G(z)$ is pole free for $R_G < |z| < \sqrt{R_G}$, or, in its weak version, for 
$R_G < |z| < 1/\sqrt{q}$.
The motivation of this hypothesis comes from the consideration of the substitution $\zeta(z := (R_G)^{s})$.
Then, we wish for a ``pole-free'' region inside the ``critical strip'' $0 \le \re(s) \le 1$, or at least, in short of general statements of a functional equation for irregular graphs, inside the open half-strip $\frac12 < \re(s) < 1$.
It turns out that there is a paucity of graph which satisfy this hypothesis, whence the weakened version stated above.
In the case of regular graphs, where $R_G = q^{-1}$, the strong and weak versions are the same.

Indeed, our A-series, without the self-adjoints, are regular and Ramanujan (except for the degenerate case of $n=1$) and do indeed obey the strong Riemann Hypothesis (with $R_G$ as well as all the pole being trivially at $1$) while with the addition of the double self-adjoints, for $n \ge 5$ both versions of the said hypothesis are violated while below 5, the strong version is satisfied except for $n=2$, which only obeys the weak.
For the D-series above, both are violated for all $n$.

\subsubsection{The Exceptional Series}
Finally, for the E-series, we also find, in the absence of self-adjoint loops, that the zeta-function is unity because of the lack of cycles. In providing the usual entry of 4 to the matrix $A$, more interesting forms for $\zeta_G(z)^{-1}$ emerge:
\begin{equation}
\begin{array}{ll}
E_6: & -(z-1)^{14} (z+1)^{13} \left(20 z^4-36 z^3+24 z^2-8 z+1\right)^2 \left(120 z^5-176
   z^4+120 z^3-48 z^2+11 z-1\right)\\
&\\
E_7: & -(z-1)^{16} (z+1)^{15} \left(100 z^6-260 z^5+280 z^4-168 z^3+60 z^2-12 z+1\right)
   \left(2400 z^9-7840 z^8\right.\\
   & \qquad \qquad \left.
+11620 z^7-10464 z^6+6356 z^5-2704 z^4+804 z^3-160 z^2+19z-1\right)\\
&\\
E_8: & -(z-1)^{18} (z+1)^{17} \left(1200000 z^{17}-8000000 z^{16}+25186000 z^{15}-49984400
   z^{14}+\right.\\
   & \qquad \qquad \left.70319040 z^{13}-74636224 z^{12}+62013952 z^{11}-41253888 z^{10}+22261008
   z^9-\right.\\
   & \qquad \qquad \left.9801936 z^8+3521088 z^7-1025344 z^6+238784 z^5-43456 z^4\right.\\
& \qquad \qquad \left. +5952 z^3-576 z^2+35
   z-1\right)

\end{array}
\end{equation}
These turn out to not to be special enough to satisfy either the strong or weak versions of the graph Riemann Hypothesis; nevertheless we will investigate shortly some spectral properties of these and the our previously encountered examples.

\subsubsection{Spectra and Roots}
The concept of the Ramanujan graph was introduced to study the spectral properties, {\it i.e.}, the gap of eigenvalues of the adjacency matrices of regular graphs.
The fact, as discussed above, that the Ramunujan property and Riemann property of graphs are equivalent is indeed elegant.
Much work has been focused on generalizing the various bounds for the maximal eigenvalues as well as the relation to to the pole-free regions of the Ihara zeta function.
Along a parallel vein has been some recent interest, continuing again on a rather old path, in the investigation of the roots of constrained polynomials (q.~v.~\cite{Derbyshire} and especially \cite{He:2010sf} in the context of gauge theories and Calabi-Yau geometries, so too is the reader referred to \cite{He:2010jh} for interpreting the zeros in the critical strip of the classical Riemann zeta function as an eigenvalue distribution).

\begin{figure}[!h!t!b]
\centerline{
$\begin{array}{ccc}
(a)
\includegraphics[trim=0mm 0mm 0mm 0mm, clip, width=2.3in]{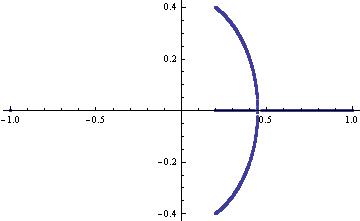}
(b)
\includegraphics[trim=0mm 0mm 0mm 0mm, clip, width=2.3in]{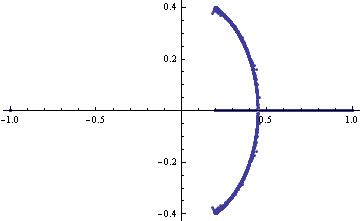}
(c)
\includegraphics[trim=0mm 0mm 0mm 0mm, clip, width=2.3in]{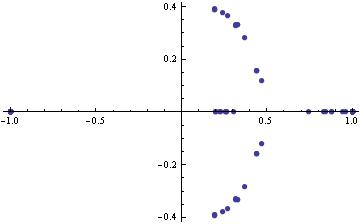}
\\
(d)
\includegraphics[trim=0mm 0mm 0mm 0mm, clip, width=2.3in]{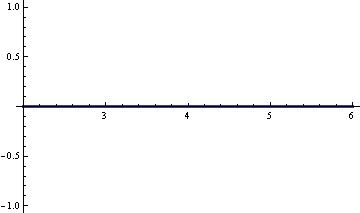}
(e)
\includegraphics[trim=0mm 0mm 0mm 0mm, clip, width=2.3in]{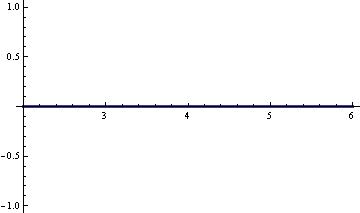}
(f)
\includegraphics[trim=0mm 0mm 0mm 0mm, clip, width=2.3in]{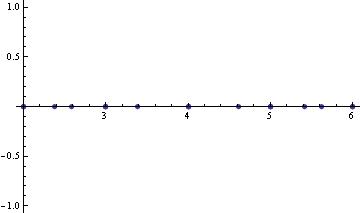}
\end{array}$
}
\caption{{\sf The distribution of the poles of the Ihara zeta function for the A-D-E quivers in (a), (b), (c) respectively, with the $A_n$ and $D_n$ series up to $n=100$, as well as the eigenvalues of their respective adjacency matrices in (d), (e) and (f).
}
\label{f:roots-ade}}
\end{figure}

It is therefore expedient to conduct some numerical experiments, similar to chapter 8 of \cite{Terras}, of the collective pole positions of the zeta function versus the spectral data of the adjacency matrices.
In \fref{f:roots-ade}, we compare and contrast these roots.
In parts (a) and (d), we present, respectively, the positions of the poles of the Ihara zeta function and the spectrum of the undirected adjacency matrix $A$ (with, as always henceforth for the ADE cases, the self-adjoining loops), for the first 100 $n$ in the $A_n$-series.
The plot (d) is rather simple because here $A$ is symmetric, whereby providing only real eigenvalues.
In (a), there are two distinctive pole-free regions: one on the real axis between $-1$ and $1/5$ (the points $\pm1$, coming from the numerator contribution in \eqref{ihara} are automatically poles) and the other, the missing conjugate pair of arcs from the semi-circle, which does not seem spurious since increasing $n$ does not appear to begin closing this gap.
In parts (b) and (e), we do the same for the D-series, up to $n=100$.
Again, the symmetry of the adjacency matrix forces real eigenvalues and we densely cover the region $[0,4]$. In (b) a hirsute decoration is added to the semi-circular distribution of poles but again the behaviour on the real line is as in the case of (a).
Finally, in (c) and (f), we plot the same for the exceptional series of $E_{6,7,8}$.

As a parting indulgence before we move to other classes of gauge theories, let us amuse ourselves with a comparative study.
Seeing the forms of the hypersurfaces in \eqref{ade}, one could not help but be reminiscent of the local Hasse-Weil zeta function, which, we recall, is the exponentiated generating function for the number of $\IF_{p^r}$-rational points of a variety over the finite number field of characteristic $q=p^r$ for some prime $p$. Recently, these functions have been computed for a host of gauge theoretic geometries in \cite{He:2010mh}.

Of course, because we are dealing with affine singular varieties, rather than smooth projective ones, immediate application of the powers of the Weil Conjectures is not at hand. However, for the A-series, having the form $xy = z^n$, the answer is simple.
For the degenerate case of $n=0$, or $\IC^2$, it is a standard fact that the Hasse-Weil local zeta function is $\zeta_p(t; \IC^2) = (1-p^2t)^{-1}$.
In general, for arbitrary $n$, we have that at $z \ne 0$, for every non-zero $x$ there is a unique $y$, giving us $(q-1)^2$ points and for $z=0$, the equation becomes $0=xy$ so that there are $2q-1$ points. In total, there are $q^2$ points over $\IF_q$ so that the local zeta function remains as $(1-p^2t)^{-1}$, amusingly reminiscent of the Ihara zeta function at $n=0$.
Since each of our graphs has an underlying affine geometry, it is perhaps fruitful to compare the Ihara and Hasse-Weil zeta functions systematically, a task we will leave for the future.

\subsection{A Zoo of Chiral Quivers}\label{s:chiral}
The ADE theories which we studied in some detail above is only a very small sample of a ``zoo'' of quiver gauge theories arising from geometry, string theory and particle physics.
Indeed, due to the fact that they are non-chiral (in that each arrow has a partner in the opposite direction between the same two nodes), interests in them have been more shown in their beautiful mathematical structure rather than their phenomenological importance.
There is a vast number of chiral theories such as (supersymmetric) Yang-Mills theories or Chern-Simons theories in various dimensions which have product gauge groups and bi-fundamental matter, and thereby affording quiver description.
Many of these have interesting vacuum moduli space prescribed by non-trivial algebraic varieties, so too do many emerge as world-volume theories of branes in string theory (cf.~\cite{Douglas:1996sw}).
Systematic and step-wise enumeration of these have only recently been undertaken \cite{Hewlett:2009bx,Davey:2009bp,Hanany:2008gx}.

It is therefore only natural that, familiarized by the ADE examples, we move to study the plethora of these chiral quiver theories.
The astute reader would point out that this is perhaps less meaningful than what was addressed previously, since in the ADE case the graph alone determined the physics since the interactions, because of the large amount of supersymmetry, can be fixed by the matter content, whereas in general we need quite more additional information. In $\cN=1$ chiral gauge theories, for example, we will need to insert a superpotential which in turn imposes an algebra on the arrows, or in the case of Chern-Simons quivers, we need levels as well as couplings.
We will return to this issue in the ensuing section which consists of an infinite family of theories completely determined by graphical data; for now, as much as to indulge our curiosity as for the sake completeness, let us move onto some representative examples of chiral $\cN=1$ quivers in 4-dimensions, which generically have Calabi-Yau threefold moduli spaces and which have potential phenomenological significance.

Therefore this brief section will only concern with the quiver part of the physics. Let us begin, with a graph quite similar to part (b) of \fref{f:eg-path}, but with 3 uni-directional arrows from one node to the next, cyclically; this is the graph for $\IC^3 / \IZ_3$ (otherwise known as del Pezzo 0) and can be thought of as the McKay quiver for $\IZ_3$ in dimension three.
Explicitly, it has $P=A={\tiny \left(
\begin{array}{ccc}
 0 & 3 & 0 \\
 0 & 0 & 3 \\
 3 & 0 & 0
\end{array}
\right)}$ and $Q = \diag(\{-1,-1,-1\})$, giving us $\zeta(z)^{-1} = 1 - 27z^3$.
The poles are situated at one-third the three cubic roots of unity, therefore the radius of converge is $R_G = 1/3$ with the entire complex plane being pole-free except on the circle of this radius, so indeed it satisfies the strong Riemann Hypothesis.

Other famous 4-dimensional theories include the canonical example of D3-branes on $\IC^3$, which is a so-called ``clover quiver'', the precursor to AdS/CFT, or the conifold theory which is a $SU(N) \times SU(N)$ theory whose classical mesonic moduli space is the affine Calabi-Yau singularity $uv-zw=0$.
In our convention, because we replace all bi-directional arrows as well as self-adjoints with a edge, the clover has a single node with $A=\{\{6\}\}, P = \{\{0\}\}$ and $Q = \{\{5\}\}$, so that $\zeta_G(z)^{-1} = (1-z^2)^2 (1-6z+5z^2)$, the same as the $A_0$ result in the ADE case.
The conifold has $A=
{\tiny \left(
\begin{array}{cc}
0 & 2 \\ 2 & 0 
\end{array}
\right)
}$, $P=0$ and $Q = \diag(\{1,1\})$, giving us $\zeta(z)^{-1} = (1 - z^2)^2$.

As we pointed out for amusement that the first example we ever gave, in part (a) of \fref{f:eg-path}, was quite akin to something well-known: the so-called Hirzebruch quivers.
These are gauge theories arising from D3-branes transverse to an affine Calabi-Yau threefold which is a complex cone over the first Hirzebruch surface, or, $\IF_0 = \IP^1 \times \IP^1$.
We have used the plural because, as it turns out, there are two well-known Seiberg dual phases (cf.~p33 of \cite{He:2004rn}), both of which are fully directed, with, respectively, $P_{I}=A_{I}= {\tiny \left(
\begin{array}{cccc}
 0 & 2 & 0 & 0 \\
 0 & 0 & 2 & 0 \\
 0 & 0 & 0 & 2 \\
 2 & 0 & 0 & 0
\end{array}
\right)}, Q_{I} = \diag(-1,-1,-1,-1)$ and
$P_{II} = A_{II} = {\tiny \left(
\begin{array}{cccc}
 0 & 2 & 0 & 2 \\
 0 & 0 & 2 & 0 \\
 4 & 0 & 0 & 0 \\
 0 & 0 & 2 & 0
\end{array}
\right),  Q_{II} = \diag(-1,-1,-1,-1)
}$.
Whence, we have $\zeta_I(z) = (1 - 16z^4)^{-1}$ and $\zeta_{II}(z) = (1-32z^3)^{-1}$ (indeed both conform to the theorem of Mizuno-Sato \cite{M-S} that for fully directed graphs, $\zeta(z) = \det(I - z A)^{-1}$) and both have poles only on some circle and thus obey the strong Riemann Hypothesis.

Now, let us investigate some partially directed graphs in our present sense.
From the taxonomy of \cite{Hanany:2008gx}, we take a few examples of quivers which arise in the study of world-volume Yang-Mills theory of D3-branes, or the Chern-Simons theory of M2-branes.
We have chosen an illustrative admixture of partially directed graphs, having various nodes, or combinations of self-adjoining loops and bi-directional arrows represented by edges.
The quiver in (a) has been affectionately dubbed the ``phallus'' and has recently been of interest in the study of M2-branes.
We see that the relevant matrices are
$A={\tiny \left(
\begin{array}{cc}
 0 & 1 \\
 1 & 4
\end{array}
\right)}$, $P=0$ and $Q=\diag(\{0,4\})$, giving us $\zeta(z)^{-1} = (1-z^2)(1-4z+3z^2)$.
The theory in (b) is the graph for the so-called ``suspended pinched point'' (SPP), an affine Calabi-Yau threefold variety described by $xy=zw^2$; it is a theory which dates back to the early days of D-brane technology.
Here we have, as a slight modification of the Cayley for $\IZ_3$, 
$A={\tiny \left(
\begin{array}{ccc}
 2 & 1 & 1 \\
 1 & 0 & 1 \\
 1 & 1 & 0
\end{array}
\right)}$, $P=0$ and $Q=\diag(\{3,1,1\})$.
This gives us $\zeta(z)^{-1}= -(z-1)^2 \left(3 z^6+4 z^5+4 z^4+2 z^3-1\right)$.
Finally, the quiver in (c) has $A = {\tiny \left(
\begin{array}{cccc}
 0 & 1 & 0 & 0 \\
 0 & 0 & 1 & 0 \\
 0 & 0 & 0 & 2 \\
 1 & 0 & 1 & 0
\end{array}
\right)}$, $P={\tiny \left(
\begin{array}{cccc}
 0 & 1 & 0 & 0 \\
 0 & 0 & 1 & 0 \\
 0 & 0 & 0 & 1 \\
 1 & 0 & 0 & 0
\end{array}
\right)}$ and $Q = \diag(\{-1,-1,0,0\})$.
This gives us
$\zeta(z) = (1-z^2-2z^4+z^6)^{-1}$.
All the three Ihara zeta functions in this set of examples are seen to satisfy the strong Riemann Hypothesis.

\begin{figure}[!h!t!b]
\centerline{
\includegraphics[trim=0mm 0mm 0mm 0mm, clip, width=6in]{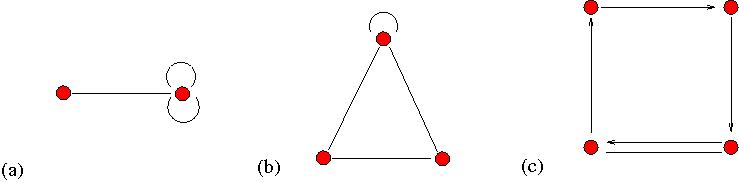}
}
\caption{{\sf Some samples of quivers which have arisen in the study of D-brane gauge theories and M-brane Chern-Simons theories.
These are all partially directed graphs, having either self-adjoining loops or bi-directional arrows represented by edges.
}
\label{f:eg-N1}}
\end{figure}

\section{Dimer Models}\label{s:dimer}
We raised the objection in the previous section, after leaving the comfortable terrain of ADE-quivers, that an immediate trouble with the zoo of quivers arising from general Yang-Mills or Chern-Simons theories is the insufficiency of the graph to completely specify the physics and the extra information requisite seem to lie beyond the scope of the zeta function.
Nevertheless, there is a vast number of tremendously interesting theories which have of late generated much beautiful mathematics and physics alike (cf,~reviews in \cite{Kennaway:2007tq,Yamazaki:2008bt}).
These are completely specified by a finite graph, {\it viz.}, a bi-partite graph (a dimer) drawn on an torus, or, equivalently, a periodic brane tiling of the plane, and has and underlying affine toric variety as the vacuum moduli space \cite{Feng:2000mi}.

We will not delve into the immensity of the subject here but will recall some key definitions first.
A {\it bi-partite graph} is one whose set of nodes can be partitions into two disjoint sets $B$ and $W$ such that no node in $B$ is adjacent to any other nodes in $B$ and likewise for $W$.
Graphically, we may represent the nodes in $B$ as, say, ``black'', and those in $W$, as ``white'', so that our dichromatic graph will have edges only between white and black nodes.

Consider such a graph as doubly periodically extended on a plane ({\it i.e.}, on a torus), with a finite representative in the fundamental region $F$.
Letting each face formed by loops in the edges in $F$ represent a gauge group factor, each edge, a matter field, such that a black (respectively white) node of degree\footnote{Or sometimes called {\bf valency}.
} $k$ contributes a degree $k$ monomial term with a plus (respectively minus) sign. Then our bi-partite graph represents a quiver gauge theory, a construction which has come to be dubbed dimer model or brane tiling. 
It is an elegant and important fact that the majority of (superconformal) gauge theories known to the AdS/CFT Correspondence fall into this class, being thus representable.

For the sake of computing the zeta function, we will not need anything as refined as the Kasteleyn matrix, which is a weighted adjacency matrix who determinant counts the number of perfect dimer matchings as well as the being intimately related to the Newton polynomial of the underlying toric variety which describes the vacuum moduli space of the gauge theory \cite{Kennaway:2007tq,Yamazaki:2008bt,Feng:2005gw}. The matrices in \eqref{ihara} shall suffice for our present purposes.

\subsection{Tilings, Bi-Partite Graphs and Ihara Zeta}
To the beautifully compiled Appendix A.1 of \cite{Davey:2009bp} we shall henceforth refer the reader, for the notation, the corresponding gauge theory and, especially, the periodic tiling by the bi-partite graph.
The programme, undertaken in the aforewritten, of extracting the various adjacency and degree matrices and computing the Ihara zeta functions, lends itself easily to us.
Indeed, because the edges in the dimer have no preferred direction, in the following examples all our $P$ matrices will be zero.

\begin{figure}[!h!t!b]
\centerline{
\includegraphics[trim=0mm 0mm 0mm 0mm, clip, width=5.5in]{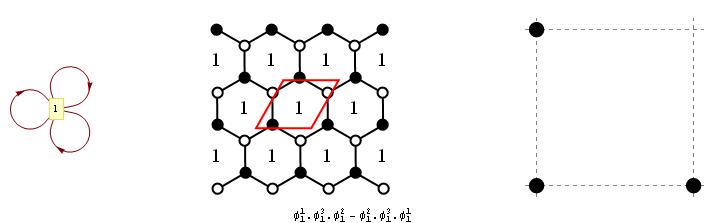}
}
\caption{{\sf 
The dimer model for $\cN=4$ super-Yang-Mills.
The three figures represent, from left to right, (a) the quiver diagram for the matter content, the single node corresponding to the $U(N)$ gauge group and the three self-adjoint loops, the three adjoint fields; (b) the hexagonal periodic bi-partite graph tiling of the plane and (c) the toric diagram of $\IC^3$, the underlying Calabi-Yau geometry. For completeness, the superpotential is also included underneath.  
The figure is drawn by the marvelous Mathematica package written by J. Pasukonis.
}
\label{f:c3dimer}}
\end{figure}

Let us commence with the simplest tiling, corresponding to the ``clover'' quiver of the $\cN=4$ Yang-Mills which launched AdS/CFT and whose vacuum moduli space is the trivial affine Calabi-Yau threefold $\IC^3$.
The dimer is a hexagonal periodic tiling, with one pair of black/white nodes, each of undirected degree (valency) 3, given in (1.1) of A.1 in {\it cit.~ibid.}, which, for illustration, we include in \fref{f:c3dimer}.
Hence, $A={\tiny \left(
\begin{array}{cc}
 0 & 3 \\
 3 & 0
\end{array}
\right)}$ and $Q = \diag(\{2,2\})$, 
giving us $\zeta(z)^{-1} = (1-z^2)(1-5z+4z^2)$.
Note that the size of these matrices is twice that of our familiar Kasteleyn matrices, this is simply because we are not addressing issues such as perfect matching and are concerned only with the adjacency information; hence a pair of black-white nodes requires dimension 2 matrices rather than 1.

It is easy to see, that in all our cases, $A = K \otimes {\tiny \left(
\begin{array}{cc}
 0 & 1 \\
 1 & 0
\end{array}
\right)}$ with $K$ the unweighted Kasteleyn matrix because we, by our convention, use the absolute adjacency and not only with the rows indexing one colour and the columns, the other.
Here, $\otimes$ is the standard Kronecker tensor product of matrices.
Similarly, $Q = R \otimes {\tiny \left(
\begin{array}{cc}
 1 & 0 \\
 0 & 1
\end{array}
\right)} - I$ , where $R$ is the (undirected) degree matrix encoding the valency of the nodes.
The reader is referred to the interesting paper \cite{Jejjala:2010vb} which has recently cast $R$ in terms of ramification indices of Groethendieck's {\it dessin d'enfants} in mapping, in the sense of Belyi, the torus on which the bi-partite graph is drawn, to a $\IP^1$.
In fact, further simplifications ensue.
Because of the so-called ``toric condition'' dubbed in \cite{Feng:2000mi}, each matter field appears precisely twice in the superpotential, with opposite sign, our bi-partite graph has the further property that no white (black) node is adjacent to more than one black (white) node.
Thus the matrix $K$ is actually diagonal, and is simply $R$.
In summary, for all our dimer models, say of $n$ pairs of black-white nodes,
\begin{equation}\label{APQdimer}
P = 0 \ ,
A = R \otimes {\tiny \left(
\begin{array}{cc}
 0 & 1 \\
 1 & 0
\end{array}
\right)} \ ,
Q = (R - I_n) \otimes {\tiny \left(
\begin{array}{cc}
 1 & 0 \\
 0 & 1
\end{array}
\right)} \ ,
\end{equation}
where $R = \diag(r_i)$ has, as its diagonal entries, the degrees of the nodes.

Aided by \eqref{APQdimer}, we move on.
Next, the conifold theory is a square tiling, with one pair of nodes so that
$A={\tiny \left(
\begin{array}{cc}
 0 & 4 \\
 4 & 0
\end{array}
\right)}$ and $Q = \diag(\{3,3\})$, yielding $\zeta(z)^{-1} = \left(1-z^2\right)^2 \left(9 z^4-10 z^2+1\right)$.
Returning to the subject of $(q+1)$-regular graphs, it is a theorem, following Perron-Frobenius, that $-(q+1)$ is in the spectrum of the adjacency matrix $A$ iff the graph is bi-partite.
Out above two examples are, respectively, bi-partite $3$- and $4$-regular graphs, and indeed $-3$ and $-4$ are eigenvalues.

Moving on to $(\IC^2 / \IZ_2) \times \IC$, presented in (2.1) of Appendix A.2, {\it Loc.~Cit.}.
The quiver for this is actually that of the $A_1$ singularity addressed earlier; however, for the dimer, we have two pairs of nodes such that 
$A={\tiny \left(
\begin{array}{cccc}
0 & 0 & 3 & 0 \\
0 & 0 & 0 & 3 \\
3 & 0 & 0 & 0 \\
0 & 3 & 0 & 0 
\end{array}
\right)}$ and $Q=\diag(\{2,2,2,2\})$, and thus $\zeta(z)^{-1} = \left(1-z^2\right)^2 \left(16 z^8-40 z^6+33 z^4-10 z^2+1\right)$.

The next example is the SPP theory by which we casually passed in \sref{s:chiral}; now let us study it more systematically.
The bi-partite graph is given in (2.2) of their Appendix A.2, with two pairs of nodes such that $A={\tiny
\left(
\begin{array}{cccc}
 0 & 0 & 3 & 0 \\
 0 & 0 & 0 & 4 \\
 3 & 0 & 0 & 0 \\
 0 & 4 & 0 & 0
\end{array}
\right)}$ and $Q = \diag(\{2,3,2,3\})$, so that $\zeta(z)^{-1} = \left(1-z^2\right)^3 \left(36 z^8-85 z^6+63 z^4-15 z^2+1\right)$.
Amidst these above examples, the Ihara zeta functions all satisfy the strong Riemann Hypothesis, except SPP, which violates even the weak version.

All periodic tilings by bi-partite graphs up to 8 pairs of black/white nodes, which are consistent in the sense of being well-defined gauge theories, have been classified in the lovely work of \cite{Davey:2009bp} and a computer database thereof, compiled.
Our examples above comprise but an hors d'oevres to this delicious database which, perhaps surprisingly - in contrast to the expectation of a myriad of contingencies - boasts only 41 consistent, inequivalent members.
This number is small enough to afford us presenting them all in Appendix \ref{ap:the41}.
In it, we present the quiver and tiling data and compute the zeta function associated to these finite graphs, indicating, in particular, whether strong/weak versions of the graph theoretic Riemann Hypothesis are obeyed or violated.
For amusement, as in \S\ref{s:chiral}, since we are naturally confronted, in addition to our bi-partite graphs, with a set of (in general) chiral quivers, it is immediate that we should compute the zeta function for these graphs as well, a simple exercise we do undertake and present together with our aforementioned appendix.

\begin{figure}[!h!t!b]
\centerline{
(a)
\includegraphics[trim=0mm 0mm 0mm 0mm,clip,width=2in]{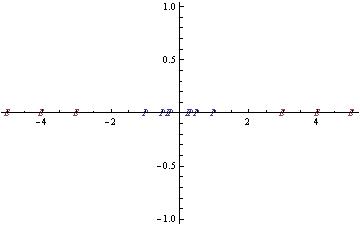}
(b)
\includegraphics[trim=0mm 0mm 0mm 0mm,clip,width=2.5in]{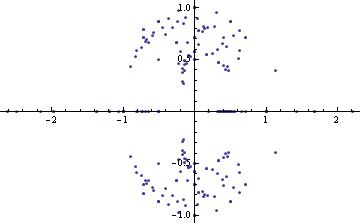}
(c)
\includegraphics[trim=0mm 0mm 0mm 0mm, clip, width=2.5in]{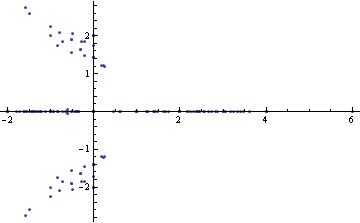}
}
\caption{{\sf 
(a) The combined plot of the poles of the Ihara zeta function (marked ``P'') and the spectrum (marked ``E'') of the adjacency matrix for the bi-partite graphs corresponding to the 41 consistent dimer models classified in the literature.
(b) The zeta poles and (c) the spectrum of the adjacency matrix of the associated (generically chiral) quivers.
}
\label{f:dimer}}
\end{figure}

As was discussed in \S\ref{s:ade}, a comparative study of the pole-free region and the spectrum of the full adjacency matrix is illuminating.
We perform this task and give the results in Figure \ref{f:dimer}.
In part (a), a compounded plot of the poles of the Ihara zeta function (marked as ``P'') and the spectrum (marked as ``E'') of the adjacency matrix (double the size of the unweighted Kasteleyn matrix) is drawn; to the simplistic appearance of the plot we will shortly return to remark.
In part (b) and (c), we respectively conglomerate the poles of the zeta functions and the eigenvalues of the adjacency matrices for all the quivers associated to the 41 consistent tilings.

The simplicity of part (a) of the figure is of no surprise to us.
From \eqref{ihara} and \eqref{APQdimer}, and recalling that $-\tr(Q-I)/2$ is equal to the difference between the number of nodes and edges, we can readily obtain the analytic expression of the zeta function.
Letting the list of valencies (ramifications) of the dimer be $\{r_1, \ldots, r_n \}$, the difference between nodes and edges is simply $2n - \sum\limits_{i=1}^n r_i$, which given that each $r_i$ is at least unity, will always be less than 0, whereby bringing the numerator down to the denominator and ensuring that the zeta function remains the reciprocal of a polynomial, as required.
Now, because the denominator has $P=0$, it reduces to 
$\det \left[
\diag\left( 1 + (r_1 - 1) z^2, \ldots, 1 + (r_n - 1) z^2 \right)
\otimes
\{ r_1, r_2, \ldots, r_n \}
\otimes
{\tiny
\left(
\begin{array}{cc}
0 & 1 \\ 1 & 0
\end{array}
\right)
}
\right]
$, giving us, upon expanding the determinant of the block matrix,
\begin{equation}
\zeta(z)^{-1} = (1-z^2)^{-2n + \sum\limits_{i=1}^n r_i}
\prod\limits_{i=1}^r
(1 + (r_i -1)z^2)^2 - (r_i z)^2 \ .
\end{equation}
Thus, the roots are at $\pm 1$ and $\pm (1-r_i)^{-1}$, all real and between $-1$ and $1$, as indicated by the figure.
Now, the radius of convergence is the distance of the closest pole to the origin, thus, $R_G = (r_{max}-1)^{-1}$, the integers $p,q$ in \eqref{pq} are $p = r_{min}-1$ and $q=r_{max}-1$, so indeed $R_G \in [q^{-1}, p^{-1}]$ and saturating the lower bound, so that the strong and weak versions of the Riemann Hypothesis coincide, which here state that the region
$(r_{max} - 1)^{-1} < |z| < (r_{max} - 1)^{-\frac12}$ is pole free in the complex plane.
In other words, all $r_i$ must be that they must satisfy the inequality: either $(r_i-1)^{-1} > (r_{max}-1)^{-\frac12}$ or $(r_i-1)^{-1} < (r_{max}-1)^{-1}$.
The second inequality being vacuous, we are thus lead to the conclusion that a
dimer model/brane-tiling satisfies the Riemann Hypothesis (as mentioned above, here the strong and weak versions equate) iff
\begin{equation}
r_i^2 - 2r_i + 2 < r_{max} \ ,
\end{equation}
for all valencies (ramifications) $r_i$ less than the maximum.

It is entertaining to meddle with a little statistics.
In \cite{Davey:2009bp}, the state of the art classification of dimer models was carried, including even the inconsistent ones in a stringent sense that they have good IR behaviour.
Nevertheless, relaxing this condition, we are granted a database of 375 graphs into which one could liberally dip. Of these, it is found that 112, or about $40.1\%$ obey the Riemann Hypothesis.
We also repeat Figure \ref{f:dimer} for these members, and not merely the distinguished 41.

\begin{figure}[!h!t!b]
\centerline{
(a)
\includegraphics[trim=0mm 0mm 0mm 0mm,clip,width=2in]{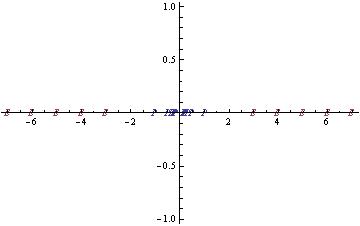}
(b)
\includegraphics[trim=0mm 0mm 0mm 0mm,clip,width=2.5in]{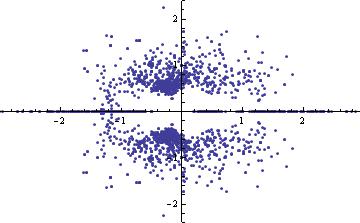}
(c)
\includegraphics[trim=0mm 0mm 0mm 0mm, clip, width=2.5in]{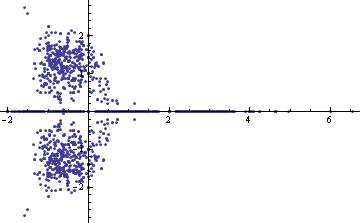}
}
\caption{{\sf 
(a) The combined plot of the poles of the Ihara zeta function (marked ``P'') and the spectrum (marked ``E'') of the adjacency matrix for the bi-partite graphs corresponding to all the known dimers, up to valency 8 and not necessarily consistent gauge theories.
(b) The zeta poles and (c) the spectrum of the adjacency matrix of the associated (generically chiral) quivers.
}
\label{f:dimer-full}}
\end{figure}

\section{Conclusions and Prospectus}\label{s:conc}
We have taken an adventurous journey accompanied by two seemingly distant acquaintances, one belonging to the world of supersymmetric gauge theories with continuous vacua and the other, to zeta functions with analogues of functional equations and zero-free regions.
The common language is that of finite graphs, either in the form of quiver representations or in the guise of periodic bi-partite tilings of the plane.

Guided by some recent explicit formulae which generalize the classical result of Ihara and which are applicable to any partially directed graph containing bi-directional arrows, cycles or self-adjoining loops, our journey is inevitably driven by Wanderlust and experimental urge.
Given the extensive catalogue, compiled over the past decade, of a myriad of gauge theories with various supersymmetry and underlying algebraic geometry, we ventured to compute the Ihara zeta-function for a host of examples.

We began by studying theories in $3+1$-dimensions with $\cN=2$ supersymmetry, whose moduli spaces are the local K3 surfaces, {\it i.e.,} quotient singularities of $\IC^2$. The graphs for these are the famous ADE quivers of McKay with two additional self-adjoining loops per node.
We find the curious fact that only the A-series for $n < 5$ satisfy the graph theoretical Riemann Hypothesis (with $n=1$ satisfying merely the weak and the others, the strong), whilst all the other quivers do not have interesting pole-free regions and violate the conditions requisite for the hypothesis.

We proceeded to the zoo of quivers of less supersymmetry, computed the Ihara zeta functions and studied their properties.
Then, anticipating the reader's apprehension that in these cases the quiver alone does not capture the physical data and that formal relations coming from the superpotential need to be imposed on the arrows, we focused primarily on the theories whose vacua are affine toric varieties.
Not only are such theories central to the AdS/CFT correspondence, they are also completely depicted by a bi-partite periodic tiling of the plane.
We investigated the Ihara zeta function for these dimer graphs and showed how to express the Graph Riemann Hypothesis in terms of inequalities in the valency numbers (order of monomials in the superpotential) of the black and white nodes.
In due course, we also examined, as inspired by Ramanujan graphs, the eigen-spectrum of the full adjacency matrices as collected roots of constrained polynomials.

It is our belief and hope that our excursion had not been a mere self-indulgence but, rather, hinted at an amusing diversion on gauge theories, leaving us with the after-taste of many a question.
What physical quantity does the Ihara zeta function measure or represent?
How do theories whose underlying graphs satisfy the strong or weak versions of the graph Riemann Hypothesis distinguish themselves?
What structure to the space of conglomerate eigenvalues of the adjacency for the set of consistent quiver gauge theories could one uncover?
To these and other enquiries let us turn, but on another day.

\newpage

\section*{Acknowledgements}
{\it Ad Catharinae Sanctae Alexandriae et ad Maiorem Dei Gloriam.}
We would like to extend our gratitude to A.~Lauder and J.~Pasukonis for valuable discussions and further to the gracious patronage of the Science and Technology Facilities Council, UK, for an Advanced Fellowship, the Chinese Ministry of Education, for a Chang-Jiang Chair Professorship at NanKai University, the National Science Foundation, USA, for grant CCF-1048082, as well as City University, London and Merton College, Oxford, for their enduring support.
Above all, to the happy occasion of the author's engagement to the lovely Miss E.~K.~Hunter this humble note is dedicated.

\appendix
\section{Ihara Zeta Functions for Brane Tilings}\label{ap:the41}
In this appendix, we catalogue the Ihara zeta functions for the 41 consistent brane-tilings (dimer model) classified in the literature \cite{Davey:2009bp}, each of which is a bi-partite graph drawn on a torus, and to which is associated a quiver gauge theory of $\cN=1$ supersymmetry and living in $3+1$-dimensions (generalization to $2+1$-dimensional Chern-Simons theories in also straight-forward).

The first column is the full adjacency matrix of the quiver diagram for the theory.
The second column records the list of the valencies or in/out degrees, {\it i.e.}, the number of edges emanating from the black/white nodes.
These two matrices completely specify the tiling and the reader is referred to \cite{Davey:2009bp} for the elegantly drawn graphs.
We then compute, in the third column, the zeta function for this information (noting that indeed, they are all reciprocals of polynomials) and check, in the fourth column, whether they obey the strong (``S''), weak (``W'') versions of the graph Riemann Hypothesis, or violate it (``N'').
For amusement and completeness, we also compute the zeta function for the quiver itself (which only partially encodes the physical data) and check their relation to the Riemann Hypothesis.

\begin{sideways}
{\scriptsize
$
\begin{array}{|l|l|l|l|l|l|}\hline
\mbox{Quiver} & \mbox{Valency} & \zeta_{Dimer}^{-1} & \mbox{RH} & 
  \zeta_{Quiver}^{-1} & \mbox{RH} \\ \hline
 \left(
\begin{array}{c}
 6
\end{array}
\right) & \{3\} & -\left(z^2-1\right)^2 \left(4 z^2-1\right) & \text{S} &
   \left(z^2-1\right)^2 \left(5 z^2-6 z+1\right) & \text{S} \\
 \left(
\begin{array}{cc}
 0 & 2 \\
 2 & 0
\end{array}
\right) & \{4\} & \left(z^2-1\right)^2 \left(9 z^4-10 z^2+1\right) & \text{S} &
   \left(z^2-1\right)^2 & \text{S} \\
 \left(
\begin{array}{cc}
 2 & 2 \\
 2 & 2
\end{array}
\right) & \{3,3\} & \left(1-4 z^2\right)^2 \left(z^2-1\right)^4 & \text{S} &
   \left(z^2-1\right)^2 \left(9 z^4-12 z^3+6 z^2-4 z+1\right) & \text{S} \\
 \left(
\begin{array}{ccc}
 0 & 1 & 1 \\
 1 & 0 & 1 \\
 1 & 1 & 2
\end{array}
\right) & \{3,4\} & -\left(z^2-1\right)^5 \left(36 z^4-13 z^2+1\right) & \text{N} & -(z-1)^2 \left(3 z^6+4 z^5+4 z^4+2 z^3-1\right) & \text{S} \\
 \left(
\begin{array}{cccc}
 0 & 0 & 0 & 2 \\
 2 & 0 & 0 & 0 \\
 0 & 2 & 0 & 0 \\
 0 & 0 & 2 & 0
\end{array}
\right) & \{4,4\} & \left(z^2-1\right)^4 \left(9 z^4-10 z^2+1\right)^2 & \text{S} &
   1-16 z^4 & \text{S} \\
 \left(
\begin{array}{cccc}
 0 & 1 & 1 & 0 \\
 1 & 0 & 0 & 1 \\
 1 & 0 & 0 & 1 \\
 0 & 1 & 1 & 0
\end{array}
\right) & \{4,4\} & \left(z^2-1\right)^4 \left(9 z^4-10 z^2+1\right)^2 & \text{S} &
   \left(z^4-1\right)^2 & \text{S} \\
 \left(
\begin{array}{ccc}
 2 & 1 & 1 \\
 1 & 2 & 1 \\
 1 & 1 & 2
\end{array}
\right) & \{3,3,3\} & -\left(z^2-1\right)^6 \left(4 z^2-1\right)^3 & \text{S} &
   -\left(z^2-1\right)^3 \left(3 z^2-z+1\right)^2 \left(3 z^2-4 z+1\right) & \text{S} \\
 \left(
\begin{array}{ccc}
 0 & 3 & 0 \\
 0 & 0 & 3 \\
 3 & 0 & 0
\end{array}
\right) & \{3,3,3\} & -\left(z^2-1\right)^6 \left(4 z^2-1\right)^3 & \text{S} & 1-27
   z^3 & \text{S} \\
 \left(
\begin{array}{cccc}
 0 & 1 & 1 & 0 \\
 1 & 0 & 0 & 1 \\
 1 & 0 & 2 & 1 \\
 0 & 1 & 1 & 2
\end{array}
\right) & \{3,3,4\} & \left(1-4 z^2\right)^2 \left(z^2-1\right)^7 \left(9 z^2-1\right)
   & \text{N} & 
\left(z^2-1\right)^2 \left(9 z^8-12 z^7+12 z^6-12 z^5+10 z^4-12 z^3+8 z^2-4 z+1\right)
& 
\text{W} \\
 \left(
\begin{array}{cccc}
 0 & 1 & 1 & 0 \\
 1 & 2 & 0 & 1 \\
 1 & 0 & 2 & 1 \\
 0 & 1 & 1 & 0
\end{array}
\right) & \{3,3,4\} & \left(1-4 z^2\right)^2 \left(z^2-1\right)^7 \left(9 z^2-1\right)
   & \text{N} & 
\left(z^2-1\right)^2 \left(9 z^8-12 z^7+16 z^6-20 z^5+14 z^4-12 z^3+8 z^2-4 z+1\right) & \text{W} \\
\hline
&&&&\mbox{{\it continued overleaf \ldots}}&\\
\end{array}
$}
\end{sideways}

\begin{sideways}
{\scriptsize
$
\begin{array}{|l|l|l|l|l|l|}\hline
\mbox{Quiver} & \mbox{Valency} & \zeta_{Dimer}^{-1} & \mbox{RH} & 
  \zeta_{Quiver}^{-1} & \mbox{RH} \\ \hline
 \left(
\begin{array}{cccc}
 0 & 0 & 0 & 2 \\
 1 & 0 & 2 & 0 \\
 1 & 0 & 0 & 1 \\
 0 & 3 & 0 & 0
\end{array}
\right) & \{3,3,4\} & \left(1-4 z^2\right)^2 \left(z^2-1\right)^7 \left(9 z^2-1\right)
   & \text{N} & -12 z^4-12 z^3+1 & \text{N} \\
 \left(
\begin{array}{ccccc}
 0 & 1 & 0 & 0 & 1 \\
 0 & 0 & 2 & 0 & 0 \\
 1 & 0 & 0 & 2 & 0 \\
 1 & 0 & 0 & 0 & 1 \\
 0 & 1 & 1 & 0 & 0
\end{array}
\right) & \{3,4,4\} & -\left(1-9 z^2\right)^2 \left(z^2-1\right)^8 \left(4
   z^2-1\right) & \text{N} & -4 z^5-12 z^4-5 z^3+1 & \text{N} \\
 \left(
\begin{array}{ccccc}
 0 & 1 & 1 & 0 & 0 \\
 1 & 0 & 0 & 1 & 0 \\
 1 & 0 & 2 & 0 & 1 \\
 0 & 1 & 0 & 0 & 1 \\
 0 & 0 & 1 & 1 & 0
\end{array}
\right) & \{3,4,4\} & -\left(1-9 z^2\right)^2 \left(z^2-1\right)^8 \left(4
   z^2-1\right) & \text{N} & -3 z^{12}+2 z^{11}+z^{10}+2 z^7-2 z^5+z^2-2 z+1
& 
\text{W} \\
 \left(
\begin{array}{cccccc}
 0 & 1 & 0 & 0 & 1 & 0 \\
 0 & 0 & 1 & 0 & 0 & 1 \\
 0 & 0 & 0 & 1 & 0 & 1 \\
 1 & 0 & 0 & 0 & 1 & 0 \\
 0 & 1 & 1 & 0 & 0 & 0 \\
 1 & 0 & 0 & 1 & 0 & 0
\end{array}
\right) & \{4,4,4\} & \left(z^2-1\right)^6 \left(9 z^4-10 z^2+1\right)^3 & \text{S} &
   -6 z^5-9 z^4-2 z^3+1 & \text{N} \\
 \left(
\begin{array}{cccccc}
 0 & 1 & 0 & 0 & 0 & 1 \\
 1 & 0 & 0 & 1 & 0 & 0 \\
 0 & 0 & 0 & 0 & 1 & 1 \\
 0 & 1 & 0 & 0 & 1 & 0 \\
 0 & 0 & 1 & 1 & 0 & 0 \\
 1 & 0 & 1 & 0 & 0 & 0
\end{array}
\right) & \{4,4,4\} & \left(z^2-1\right)^6 \left(9 z^4-10 z^2+1\right)^3 & \text{S} &
   \left(z^6-1\right)^2 & \text{S} \\
 \left(
\begin{array}{cccccc}
 0 & 0 & 0 & 2 & 0 & 0 \\
 0 & 0 & 0 & 0 & 2 & 0 \\
 0 & 0 & 0 & 0 & 0 & 2 \\
 0 & 1 & 1 & 0 & 0 & 0 \\
 1 & 0 & 1 & 0 & 0 & 0 \\
 1 & 1 & 0 & 0 & 0 & 0
\end{array}
\right) & \{4,4,4\} & \left(z^2-1\right)^6 \left(9 z^4-10 z^2+1\right)^3 & \text{S} &
   -16 z^6-12 z^4+1 & \text{S} \\
 \left(
\begin{array}{cccc}
 2 & 1 & 0 & 1 \\
 1 & 2 & 1 & 0 \\
 0 & 1 & 2 & 1 \\
 1 & 0 & 1 & 2
\end{array}
\right) & \{3,3,3,3\} & \left(1-4 z^2\right)^4 \left(z^2-1\right)^8 & \text{S} 
&
\left(z^2-1\right)^4 \left(3 z^2-2 z+1\right)^2 \left(9 z^4-12 z^3+6 z^2-4 z+1\right)
   & \text{S} \\
\hline
&&&&\mbox{{\it continued overleaf \ldots}}&\\
\end{array}
$}
\end{sideways}

\begin{sideways}
{\scriptsize
$
\begin{array}{|l|l|l|l|l|l|}\hline
\mbox{Quiver} & \mbox{Valency} & \zeta_{Dimer}^{-1} & \mbox{RH} & 
  \zeta_{Quiver}^{-1} & \mbox{RH} \\ \hline
 \left(
\begin{array}{cccc}
 0 & 2 & 0 & 1 \\
 0 & 0 & 1 & 2 \\
 2 & 1 & 0 & 0 \\
 1 & 0 & 2 & 0
\end{array}
\right) & \{3,3,3,3\} & \left(1-4 z^2\right)^4 \left(z^2-1\right)^8 & \text{S} & -16
   z^8+32 z^6-16 z^4-16 z^3+1 & \text{N} \\
 \left(
\begin{array}{cccc}
 0 & 0 & 0 & 2 \\
 0 & 0 & 0 & 2 \\
 2 & 2 & 0 & 0 \\
 0 & 0 & 4 & 0
\end{array}
\right) & \{3,3,3,3\} & \left(1-4 z^2\right)^4 \left(z^2-1\right)^8 & \text{S} & 1-32
   z^3 & \text{S} \\
 \left(
\begin{array}{cccc}
 0 & 1 & 1 & 1 \\
 1 & 0 & 1 & 1 \\
 1 & 1 & 0 & 1 \\
 1 & 1 & 1 & 0
\end{array}
\right) & \{3,3,3,3\} & \left(1-4 z^2\right)^4 \left(z^2-1\right)^8 & \text{S} &
   \left(z^2-1\right)^2 \left(2 z^2-3 z+1\right) \left(2 z^2+z+1\right)^3 & \text{S}
   \\
 \left(
\begin{array}{ccccc}
 0 & 1 & 1 & 0 & 0 \\
 1 & 0 & 0 & 0 & 1 \\
 1 & 0 & 2 & 1 & 0 \\
 0 & 0 & 1 & 2 & 1 \\
 0 & 1 & 0 & 1 & 2
\end{array}
\right) & \{3,3,3,4\} & -\left(z^2-1\right)^9 \left(4 z^2-1\right)^3 \left(9
   z^2-1\right) & \text{N} & 
-\left(z^2-1\right)^3 \left(27 z^{10}-54 z^9+66 z^8-70 z^7+70 z^6-72 z^5+58 z^4-38z^3+18 z^2-6 z+1\right)
& \text{W} \\
 \left(
\begin{array}{ccccc}
 0 & 1 & 1 & 0 & 0 \\
 1 & 2 & 0 & 0 & 1 \\
 1 & 0 & 2 & 1 & 0 \\
 0 & 0 & 1 & 2 & 1 \\
 0 & 1 & 0 & 1 & 0
\end{array}
\right) & \{3,3,3,4\} & -\left(z^2-1\right)^9 \left(4 z^2-1\right)^3 \left(9
   z^2-1\right) & \text{N} & 
-\left(z^2-1\right)^3 \left(27 z^{10}-54 z^9+78 z^8-102 z^7+102 z^6-88 z^5+62 z^4-38
   z^3+18 z^2-6 z+1\right)
& \text{W} \\
 \left(
\begin{array}{ccccc}
 0 & 1 & 0 & 0 & 1 \\
 0 & 0 & 3 & 0 & 0 \\
 2 & 0 & 0 & 2 & 0 \\
 0 & 1 & 0 & 0 & 1 \\
 0 & 1 & 1 & 0 & 0
\end{array}
\right) & \{3,3,3,4\} & -\left(z^2-1\right)^9 \left(4 z^2-1\right)^3 \left(9
   z^2-1\right) & \text{N} & -12 z^4-16 z^3+1 & \text{N} \\
 \left(
\begin{array}{ccccc}
 0 & 1 & 0 & 1 & 0 \\
 0 & 0 & 2 & 0 & 1 \\
 1 & 0 & 0 & 1 & 1 \\
 0 & 2 & 1 & 0 & 0 \\
 1 & 0 & 0 & 1 & 0
\end{array}
\right) & \{3,3,3,4\} & -\left(z^2-1\right)^9 \left(4 z^2-1\right)^3 \left(9
   z^2-1\right) & \text{N} & 4 z^7+8 z^6-5 z^5-13 z^4-11 z^3+1 & \text{N} \\
\hline
&&&&\mbox{{\it continued overleaf \ldots}}&\\
\end{array}
$}
\end{sideways}

\begin{sideways}
{\scriptsize
$
\begin{array}{|l|l|l|l|l|l|}\hline
\mbox{Quiver} & \mbox{Valency} & \zeta_{Dimer}^{-1} & \mbox{RH} & 
  \zeta_{Quiver}^{-1} & \mbox{RH} \\ \hline
 \left(
\begin{array}{cccccc}
 0 & 1 & 1 & 0 & 0 & 0 \\
 1 & 0 & 0 & 0 & 1 & 0 \\
 1 & 0 & 2 & 1 & 0 & 0 \\
 0 & 0 & 1 & 2 & 0 & 1 \\
 0 & 1 & 0 & 0 & 0 & 1 \\
 0 & 0 & 0 & 1 & 1 & 0
\end{array}
\right) & \{3,3,4,4\} & \left(z^2-1\right)^{10} \left(36 z^4-13 z^2+1\right)^2 &
   \text{N} & 
\begin{array}{c}
(z-1)^3 (z+1)^2 \left(9 z^{11}-3 z^{10}+9 z^9-3 z^8+9 z^7- \right. \\
\left. 3 z^6+7 z^5-5 z^4+7 z^3-5 z^2+3 z-1\right)
\end{array}
& \text{S} \\
 \left(
\begin{array}{cccccc}
 0 & 1 & 1 & 0 & 0 & 0 \\
 1 & 0 & 0 & 0 & 1 & 0 \\
 1 & 0 & 2 & 0 & 0 & 1 \\
 0 & 0 & 0 & 0 & 1 & 1 \\
 0 & 1 & 0 & 1 & 2 & 0 \\
 0 & 0 & 1 & 1 & 0 & 0
\end{array}
\right) & \{3,3,4,4\} & \left(z^2-1\right)^{10} \left(36 z^4-13 z^2+1\right)^2 &
   \text{N} & 
\begin{array}{c}
9 z^{16}-12 z^{15}-2 z^{14}+4 z^{13}+z^{12}-10 z^{10}+16 z^9-\\
4 z^8-2 z^6+z^4-4 z^3+6
   z^2-4 z+1
\end{array}
& \text{W} \\
 \left(
\begin{array}{cccccc}
 0 & 1 & 1 & 0 & 0 & 0 \\
 1 & 2 & 0 & 0 & 1 & 0 \\
 1 & 0 & 2 & 0 & 0 & 1 \\
 0 & 0 & 0 & 0 & 1 & 1 \\
 0 & 1 & 0 & 1 & 0 & 0 \\
 0 & 0 & 1 & 1 & 0 & 0
\end{array}
\right) & \{3,3,4,4\} & \left(z^2-1\right)^{10} \left(36 z^4-13 z^2+1\right)^2 &
   \text{N} & 
\begin{array}{c}
(z-1)^3 (z+1)^2 \left(9 z^{11}-3 z^{10}+13 z^9-7 z^8+13 z^7-\right.\\
\left. 7 z^6+11 z^5-9 z^4+7 z^3-5 z^2+3 z-1\right)
\end{array}
& \text{W} \\
 \left(
\begin{array}{cccccc}
 0 & 0 & 0 & 2 & 0 & 0 \\
 0 & 0 & 0 & 0 & 0 & 2 \\
 1 & 0 & 0 & 0 & 2 & 0 \\
 0 & 1 & 2 & 0 & 0 & 0 \\
 0 & 1 & 0 & 1 & 0 & 0 \\
 1 & 0 & 1 & 0 & 0 & 0
\end{array}
\right) & \{3,3,4,4\} & \left(z^2-1\right)^{10} \left(36 z^4-13 z^2+1\right)^2 &
   \text{N} & -16 z^6-8 z^5-8 z^4-8 z^3+1 & \text{N} \\
 \left(
\begin{array}{cccccc}
 0 & 1 & 0 & 0 & 0 & 2 \\
 0 & 0 & 1 & 0 & 1 & 0 \\
 1 & 0 & 0 & 1 & 0 & 0 \\
 1 & 0 & 0 & 0 & 0 & 1 \\
 1 & 0 & 0 & 1 & 0 & 0 \\
 0 & 1 & 1 & 0 & 1 & 0
\end{array}
\right) & \{3,3,3,5\} & \left(z^2-1\right)^{10} \left(4 z^2-1\right)^3 \left(16
   z^2-1\right) & \text{S} & -4 z^5-12 z^4-8 z^3+1 & \text{N} \\
 \left(
\begin{array}{cccccc}
 0 & 1 & 0 & 0 & 1 & 1 \\
 0 & 0 & 0 & 1 & 1 & 0 \\
 1 & 0 & 0 & 0 & 0 & 1 \\
 1 & 0 & 1 & 0 & 0 & 0 \\
 1 & 0 & 1 & 1 & 0 & 0 \\
 0 & 1 & 0 & 0 & 1 & 0
\end{array}
\right) & \{3,3,3,5\} & \left(z^2-1\right)^{10} \left(4 z^2-1\right)^3 \left(16
   z^2-1\right) & \text{S} & 4 z^7+2 z^6-6 z^5-11 z^4-6 z^3+1 & \text{N} \\
\hline
&&&&\mbox{{\it continued overleaf \ldots}}&\\
\end{array}
$}
\end{sideways}

\begin{sideways}
{\tiny
$
\begin{array}{|l|l|l|l|l|l|}\hline
\mbox{Quiver} & \mbox{Valency} & \zeta_{Dimer}^{-1} & \mbox{RH} & 
  \zeta_{Quiver}^{-1} & \mbox{RH} \\ \hline
 \left(
\begin{array}{cccccc}
 0 & 1 & 0 & 0 & 1 & 0 \\
 0 & 0 & 2 & 0 & 0 & 1 \\
 1 & 0 & 0 & 1 & 0 & 1 \\
 0 & 1 & 0 & 0 & 1 & 0 \\
 0 & 1 & 1 & 0 & 0 & 0 \\
 1 & 0 & 0 & 1 & 0 & 0
\end{array}
\right) & \{3,3,4,4\} & \left(z^2-1\right)^{10} \left(36 z^4-13 z^2+1\right)^2 &
   \text{S} & -4 z^5-12 z^4-8 z^3+1 & \text{N} \\ \hline
 \left(
\begin{array}{cccccc}
 0 & 1 & 0 & 0 & 1 & 0 \\
 0 & 0 & 1 & 0 & 0 & 1 \\
 2 & 0 & 0 & 2 & 0 & 0 \\
 0 & 1 & 0 & 0 & 1 & 0 \\
 0 & 0 & 1 & 0 & 0 & 1 \\
 0 & 0 & 2 & 0 & 0 & 0
\end{array}
\right) & \{3,3,4,4\} & \left(z^2-1\right)^{10} \left(36 z^4-13 z^2+1\right)^2 &
   \text{N} & -16 z^4-8 z^3+1 & \text{N} \\
 \left(
\begin{array}{cccccc}
 0 & 1 & 0 & 1 & 0 & 0 \\
 0 & 0 & 1 & 0 & 0 & 1 \\
 1 & 0 & 0 & 1 & 1 & 0 \\
 0 & 1 & 1 & 0 & 0 & 1 \\
 1 & 0 & 0 & 1 & 0 & 0 \\
 0 & 0 & 1 & 0 & 1 & 0
\end{array}
\right) & \{3,3,4,4\} & \left(z^2-1\right)^{10} \left(36 z^4-13 z^2+1\right)^2 &
   \text{N} & 4 z^7+2 z^6-6 z^5-11 z^4-6 z^3+1 & \text{N} \\
 \left(
\begin{array}{ccccccc}
 0 & 1 & 1 & 0 & 0 & 0 & 0 \\
 1 & 0 & 0 & 0 & 1 & 0 & 0 \\
 1 & 0 & 2 & 0 & 0 & 0 & 1 \\
 0 & 0 & 0 & 0 & 0 & 1 & 1 \\
 0 & 1 & 0 & 0 & 0 & 1 & 0 \\
 0 & 0 & 0 & 1 & 1 & 0 & 0 \\
 0 & 0 & 1 & 1 & 0 & 0 & 0
\end{array}
\right) & \{3,4,4,4\} & -\left(z^2-1\right)^{11} \left(4 z^2-1\right) \left(9
   z^2-1\right)^3 & \text{N} & -3 z^{16}+2 z^{15}+z^{14}+2 z^9-2 z^7+z^2-2 z+1
& \text{W}
   \\
 \left(
\begin{array}{ccccccc}
 0 & 0 & 0 & 1 & 1 & 0 & 0 \\
 0 & 0 & 0 & 0 & 0 & 0 & 2 \\
 0 & 0 & 0 & 0 & 0 & 2 & 0 \\
 0 & 0 & 1 & 0 & 1 & 0 & 0 \\
 0 & 1 & 0 & 0 & 0 & 1 & 0 \\
 1 & 1 & 0 & 1 & 0 & 0 & 0 \\
 1 & 0 & 1 & 0 & 0 & 0 & 0
\end{array}
\right) & \{3,4,4,4\} & -\left(z^2-1\right)^{11} \left(4 z^2-1\right) \left(9
   z^2-1\right)^3 & \text{N} & -14 z^6-4 z^5-9 z^4-4 z^3+1 & \text{N} \\
 \left(
\begin{array}{ccccccc}
 0 & 1 & 0 & 0 & 1 & 0 & 0 \\
 0 & 0 & 1 & 0 & 0 & 0 & 1 \\
 1 & 0 & 0 & 1 & 0 & 1 & 0 \\
 0 & 1 & 0 & 0 & 1 & 0 & 0 \\
 0 & 0 & 1 & 0 & 0 & 0 & 1 \\
 1 & 0 & 0 & 1 & 0 & 0 & 0 \\
 0 & 0 & 1 & 0 & 0 & 1 & 0
\end{array}
\right) & \{3,4,4,4\} & -\left(z^2-1\right)^{11} \left(4 z^2-1\right) \left(9
   z^2-1\right)^3 & \text{N} & -4 z^5-12 z^4-4 z^3+1 & \text{N} \\
 \left(
\begin{array}{cccccccc}
 0 & 1 & 0 & 0 & 0 & 0 & 0 & 1 \\
 1 & 0 & 0 & 0 & 1 & 0 & 0 & 0 \\
 0 & 0 & 0 & 0 & 0 & 0 & 1 & 1 \\
 0 & 0 & 0 & 0 & 0 & 1 & 1 & 0 \\
 0 & 1 & 0 & 0 & 0 & 1 & 0 & 0 \\
 0 & 0 & 0 & 1 & 1 & 0 & 0 & 0 \\
 0 & 0 & 1 & 1 & 0 & 0 & 0 & 0 \\
 1 & 0 & 1 & 0 & 0 & 0 & 0 & 0
\end{array}
\right) & \{4,4,4,4\} & \left(z^2-1\right)^8 \left(9 z^4-10 z^2+1\right)^4 & \text{S}
   & \left(z^8-1\right)^2 & \text{S} \\
&&&&\mbox{{\it continued overleaf \ldots}}&\\
\end{array}
$}
\end{sideways}

\begin{sideways}
{\scriptsize
$
\begin{array}{|l|l|l|l|l|l|}\hline
\mbox{Quiver} & \mbox{Valency} & \zeta_{Dimer}^{-1} & \mbox{RH} & 
  \zeta_{Quiver}^{-1} & \mbox{RH} \\ \hline
 \left(
\begin{array}{cccccccc}
 0 & 0 & 2 & 0 & 0 & 0 & 0 & 0 \\
 0 & 0 & 0 & 0 & 0 & 0 & 2 & 0 \\
 0 & 0 & 0 & 1 & 1 & 0 & 0 & 0 \\
 0 & 0 & 0 & 0 & 0 & 0 & 0 & 2 \\
 0 & 0 & 0 & 0 & 0 & 2 & 0 & 0 \\
 1 & 1 & 0 & 0 & 0 & 0 & 0 & 0 \\
 0 & 0 & 0 & 1 & 1 & 0 & 0 & 0 \\
 1 & 1 & 0 & 0 & 0 & 0 & 0 & 0
\end{array}
\right) & \{4,4,4,4\} & \left(z^2-1\right)^8 \left(9 z^4-10 z^2+1\right)^4 & \text{S}
   & 1-16 z^4 & \text{S} \\
 \left(
\begin{array}{cccccccc}
 0 & 0 & 0 & 1 & 1 & 0 & 0 & 0 \\
 0 & 0 & 0 & 0 & 0 & 0 & 2 & 0 \\
 0 & 0 & 0 & 0 & 0 & 1 & 0 & 1 \\
 0 & 0 & 1 & 0 & 1 & 0 & 0 & 0 \\
 0 & 1 & 0 & 0 & 0 & 1 & 0 & 0 \\
 1 & 0 & 0 & 0 & 0 & 0 & 0 & 1 \\
 1 & 0 & 1 & 0 & 0 & 0 & 0 & 0 \\
 0 & 1 & 0 & 1 & 0 & 0 & 0 & 0
\end{array}
\right) & \{4,4,4,4\} & \left(z^2-1\right)^8 \left(9 z^4-10 z^2+1\right)^4 & \text{S}
   & 4 z^8-4 z^7-7 z^6-4 z^5-8 z^4-2 z^3+1 & \text{N} \\
 \left(
\begin{array}{cccccccc}
 0 & 1 & 0 & 0 & 0 & 0 & 1 & 0 \\
 0 & 0 & 0 & 1 & 0 & 0 & 0 & 1 \\
 1 & 0 & 0 & 1 & 0 & 0 & 0 & 0 \\
 0 & 0 & 0 & 0 & 0 & 1 & 1 & 0 \\
 0 & 1 & 1 & 0 & 0 & 0 & 0 & 0 \\
 1 & 0 & 0 & 0 & 1 & 0 & 0 & 0 \\
 0 & 0 & 0 & 0 & 1 & 0 & 0 & 1 \\
 0 & 0 & 1 & 0 & 0 & 1 & 0 & 0
\end{array}
\right) & \{4,4,4,4\} & \left(z^2-1\right)^8 \left(9 z^4-10 z^2+1\right)^4 & \text{S}
   & -16 z^6-12 z^4+1 & \text{S} \\
 \left(
\begin{array}{cccccccc}
 0 & 1 & 0 & 0 & 1 & 0 & 0 & 0 \\
 0 & 0 & 1 & 0 & 0 & 0 & 1 & 0 \\
 0 & 0 & 0 & 0 & 0 & 1 & 0 & 1 \\
 0 & 1 & 0 & 0 & 1 & 0 & 0 & 0 \\
 0 & 0 & 1 & 0 & 0 & 0 & 1 & 0 \\
 1 & 0 & 0 & 1 & 0 & 0 & 0 & 0 \\
 0 & 0 & 0 & 0 & 0 & 1 & 0 & 1 \\
 1 & 0 & 0 & 1 & 0 & 0 & 0 & 0
\end{array}
\right) & \{4,4,4,4\} & \left(z^2-1\right)^8 \left(9 z^4-10 z^2+1\right)^4 & \text{S}
   & 1-16 z^4 & \text{S}
\\ \hline
\end{array}
$}
\end{sideways}


\end{document}